\documentclass[prl,twocolumn,showpacs,amsmath,amssymb,superscriptaddress,nofootinbib]{revtex4-1}
\usepackage[utf8]{inputenc}
\usepackage{epsfig}
\usepackage{graphicx}% Include figure files
\usepackage{dcolumn}% Align table columns on decimal point
\usepackage{bm}% bold math
\usepackage{color}
\usepackage{url}
\usepackage{ulem}
\usepackage{multirow}
\usepackage{enumitem}
\usepackage{CJK}
\usepackage[bookmarksnumbered,bookmarksopen,colorlinks,citecolor=red,linkcolor=blue]{hyperref}

\newcommand{\pT} {p_{\rm T}}

\newcommand{\bB}{{\bm B}}
\newcommand{\bE}{{\bm E}}
\newcommand{\br}{{\bm r}}
\newcommand{\bv}{{\bm v}}

\usepackage{tikz,xcolor,hyperref}
\definecolor{lime}{HTML}{A6CE39}
\DeclareRobustCommand{\orcidicon}{
	\begin{tikzpicture}
	\draw[lime, fill=lime] (0,0) 
	circle [radius=0.16] 
	node[white] {{\fontfamily{qag}\selectfont \tiny ID}};
	\draw[white, fill=white] (-0.0625,0.095) 
	circle [radius=0.007];
	\end{tikzpicture}
	\hspace{-2mm}
}

\foreach \x in {A, ..., Z}{%
	\expandafter\xdef\csname orcid\x\endcsname{\noexpand\href{https://orcid.org/\csname orcidauthor\x\endcsname}{\noexpand\orcidicon}}
}

\begin{document}
\title{Electromagnetic fields in ultra-peripheral relativistic heavy-ion collisions}\thanks{This work is supported in part by the National Key Research and Development Program of China (Nos. 2022YFA1604900), the Guangdong Major Project of Basic and Applied Basic Research (No. 2020B0301030008), the National Natural Science Foundation of China (Nos. 12275053, 12025501, 11890710, 11890714, 12147101, 12075061, and 12225502), the Strategic Priority Research Program of Chinese Academy of Sciences (No. XDB34030000), Shanghai National Science Foundation (No. 20ZR1404100), and STCSM (No. 23590780100).
}

\author{Jie Zhao\orcidA{}}
	\address{Key Laboratory of Nuclear Physics and Ion-beam Application (MOE), Institute of Modern Physics, Fudan University, Shanghai 200433, China}
	\address{Shanghai Research Center for Theoretical Nuclear Physics, NSFC and Fudan University, Shanghai 200438, China}
\author{Jin-Hui Chen\orcidB{}}
\email{chenjinhui@fudan.edu.cn}
	\address{Key Laboratory of Nuclear Physics and Ion-beam Application (MOE), Institute of Modern Physics, Fudan University, Shanghai 200433, China}
	\address{Shanghai Research Center for Theoretical Nuclear Physics, NSFC and Fudan University, Shanghai 200438, China}
\author{Xu-Guang Huang \orcidC{}}
\email{huangxuguang@fudan.edu.cn}
	\address{Physics Department and Center for Particle Physics and Field Theory, Fudan University, Shanghai 200438, China}
	\address{Key Laboratory of Nuclear Physics and Ion-beam Application (MOE), Institute of Modern Physics, Fudan University, Shanghai 200433, China}
	\address{Shanghai Research Center for Theoretical Nuclear Physics, NSFC and Fudan University, Shanghai 200438, China}
\author{Yu-Gang Ma \orcidD{}}
	\address{Key Laboratory of Nuclear Physics and Ion-beam Application (MOE), Institute of Modern Physics, Fudan University, Shanghai 200433, China}
	\address{Shanghai Research Center for Theoretical Nuclear Physics, NSFC and Fudan University, Shanghai 200438, China}

\begin{abstract}

	Ultraperipheral heavy-ion collisions (UPCs) offer unique opportunities to study processes under strong electromagnetic fields. In these collisions, highly charged fast-moving ions carry strong electromagnetic fields that can be effectively treated as photon fluxes. The exchange of photons can induce photonuclear and two-photon interactions, and excite ions. This excitation of the ions results in Coulomb dissociation with the emission of photons, neutrons, and other particles. 	Additionally, the electromagnetic fields generated by the ions can be sufficiently strong to enforce mutual interactions between the two colliding ions. Consequently, the two colliding ions experienced an electromagnetic force that pushed them in opposite directions, causing a back-to-back correlation in the emitted neutrons. Using a Monte Carlo simulation, we qualitatively demonstrated that the above electromagnetic effect is large enough to be observed in UPCs, which would provide a clear means to study strong electromagnetic fields and their effects. 

\end{abstract}
\maketitle

\section{Introduction}  
Over the past two decades, several novel phenomena associated with strong electromagnetic fields in hot quantum chromodynamics (QCD) have been proposed, such as the chiral magnetic effect (CME)~\cite{Kharzeev:1998kz,Huang:2015oca,Wang:2018ygc,Zhao:2019hta,Fang:2021cpl,Liu:2022Sci}.
Such strong electromagnetic fields are expected to be generated during relativistic heavy-ion collisions; however, they are extremely challenging to measure experimentally. Several attempts have been made to detect strong electromagnetic fields in heavy ion collisions~\cite{STAR:2017ckg,STAR:2021nst,Liu:2020nst,Gao:2020nst}. However, electromagnetic processes are obscured by strong hadronic interactions when nuclei collide. Conversely, ultraperipheral heavy-ion collisions (UPCs) offer a distinctive advantage in observing electromagnetic processes, as the impact parameters in these collisions are more than twice the nuclear radius, preventing hadronic interactions~\cite{Bertulani:2005ru,Klein:2016yzr,Klein:2020fmr}. Therefore, UPCs provide a unique opportunity to study strong electromagnetic processes in relativistic heavy ion collisions~\cite{Baur:1998ay,Bertulani:2005ru}. 

In UPCs, fast-moving heavy ions are accompanied by intense photon fluxes due to their large electric charges and strong Lorentz-contracted electromagnetic fields. These fields are sufficiently strong to induce photonuclear and photon-photon interactions~\cite{Bertulani:2005ru,Klein:2016yzr,Klein:2020fmr}. Many interesting results regarding particle production, including $e^{+}+e^{-}$ and $\pi^{+}+\pi^{-}$, and vector mesons, such as $\rho, \omega, J/\psi$ have been reported in UPCs~\cite{Klein:1999qj,STAR:2002caw,PHENIX:2009xtn,ALICE:2012yye,CMS:2016itn,ATLAS:2017fur,STAR:2019wlg}. Thus, it is natural to assume that these strong electromagnetic fields could likewise have visible impacts on collision dynamics through the electromagnetic force.

The most important region for Coulomb dissociation is the giant dipole resonance (GDRs)~\cite{Goldhaber:1948zza}, approximately a few MeVs. 
The photon-nuclear interaction in UPCs has a large cross-section to produce GDRs or excite nuclei to high-energy states~\cite{Berman:1975tt,Chomaz:1993qe}. The GDRs typically decay by emitting photons or neutrons. The excitation energy of the GDRs is approximately 10-14 MeV for heavy nuclei and higher for lighter nuclei~\cite{Berman:1975tt,Veyssiere:1970ztg,Tao:2013PRC,He:2014PRL,Huang:2021PRC}. The emitted neutrons have similar energies, which are considerably lower than the typical energy scales of relativistic heavy-ion collisions. Consequently, the emitted neutrons can be used to detect strong electromagnetic fields. In this study, we employed Monte (MC) simulations to demonstrate that the electromagnetic effect on neutron emission from Coulomb dissociation, primarily through GDRs decay, is sufficiently significant to be observed in UPCs. Our simulation incorporates the calculation of the strong electromagnetic field using Li\'enard-Wiechert potential, combined with neutron emission data from the existing experimental results and model simulations. Natural units $\hbar=c=1$ are used in this study.

\begin{figure*}[htb]
	\centering
\includegraphics[width=0.5\textwidth]{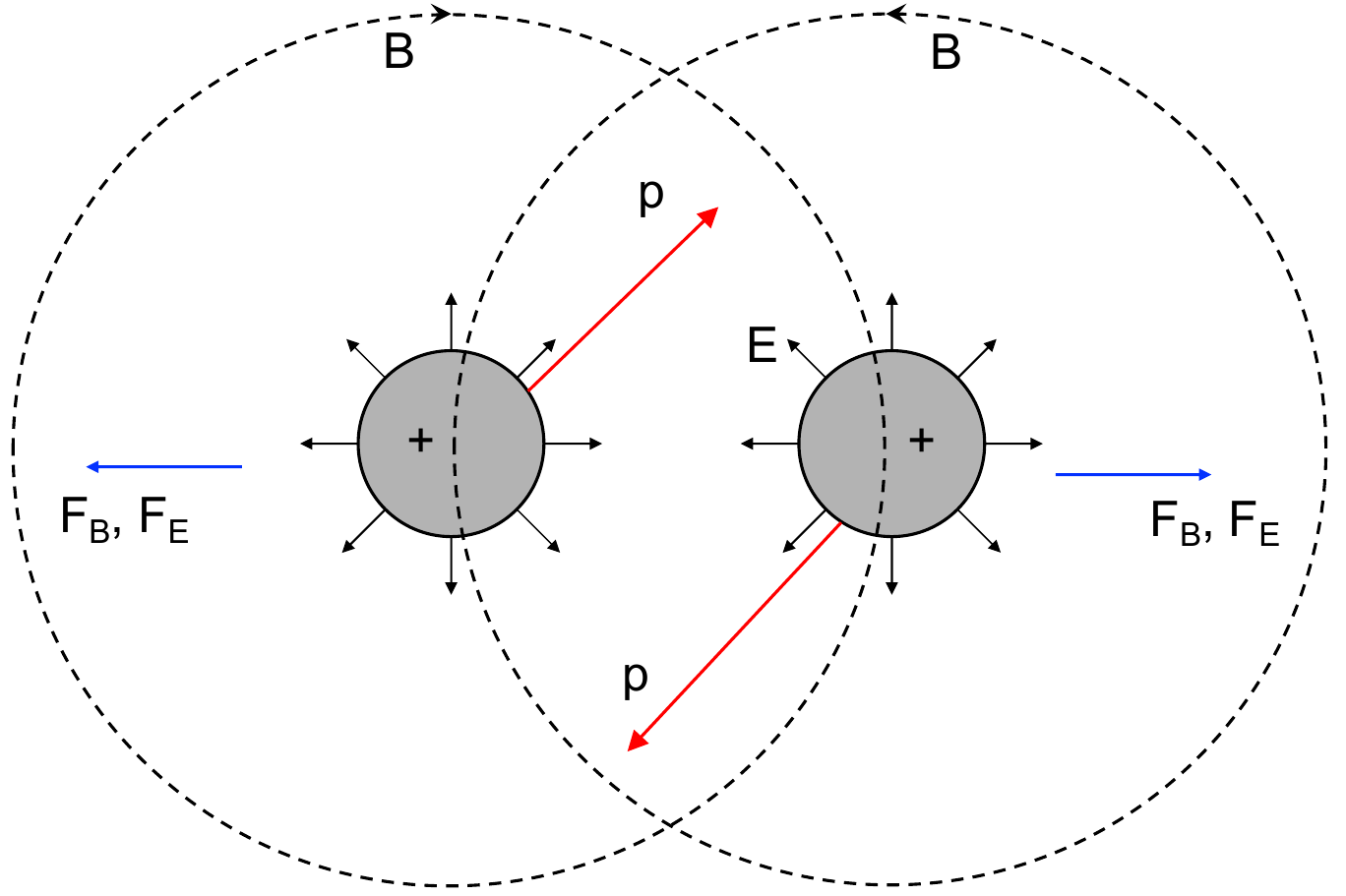}
\caption[]{(Color online) 
	Schematic view of electromagnetic forces in an ultra-peripheral relativistic heavy-ion collision, where the impact parameter $b > 2R$ ($R$ is the nuclear radius). 
	Red arrows indicate the moving directions of the two colliding nuclei. 
	The two fast-moving nuclei are highly charged (positively electric charged). The electric-field directions are indicated by the black solid arrows. 
	The black-dashed lines show the magnetic-field directions. 
	Both the electric and Lorentz forces push the two nuclei in opposite directions, as indicated by the blue arrows.
	The electromagnetic forces are strong enough to impact the emission of particles from the nuclei. For instance, the neutrons, emitted from the Coulomb dissociation of two colliding ions (such as the GDRs decay), will have a back-to-back correlation in the transverse plane.
}
\label{plotEBfile}
\end{figure*}

\section{General idea} 

When two ions collide head-on in the UPC, a strong electromagnetic field is generated. For instance, in a gold-gold (Au + Au) UPC with an impact parameter of $b=20$ ~ fm~\cite{Baltz:2002pp} and a center-of-mass energy of $\sqrt{s_\text{NN}} = 200$ GeV, as shown in Fig. ~\ref{plotEBfile}, the electric and magnetic fields produced by one nucleus acting on another are approximately $eE\sim0.2m_{\pi}^2$ and $eB\sim0.2m_{\pi}^2$, respectively, ~\cite{Deng:2012pc,Bzdak:2011yy}. Electric and Lorentz forces would push the two colliding ions in opposite directions in the transverse plane. Assuming that the nuclei have no initial transverse moment, the velocity of each nucleon can be calculated using the following equation: 
\begin{equation}
	v_{z}^{2} = 1-(2m_\text{N}/\sqrt{s_\text{NN}})^{2}, 
\end{equation}
where $m_\text{N}$ denotes nucleon mass. At the RHIC and LHC energies, where $v_z$ is very large, the nuclei are strongly Lorentz contracted and adopt a pancake-like shape. The effective collision time was estimated as $\Delta t = 2R/(\gamma v_{z})$~\cite{Bertulani:2005ru}, where $R$ is the radius of the nucleus and $\gamma=1/\sqrt{(1-v_z^2)}$ is the Lorentz factor, which is approximately 107 for Au + Au collisions at $\sqrt{s_\text{NN}} = 200$ GeV. Hence, during the collision, the transverse momentum per nucleon due to the strong electromagnetic field is approximately 2 MeV by a rough estimate [$\Delta p = Z(eE+v_z eB)\Delta t/A\approx 0.4m_{\pi}^2\times0.12\,{\rm fm}\times79/197\sim2$ MeV, with $Z = 79, A = 197$, and $R = 6.38$ fm for the Au nucleus]. 

However, the excitation energy of a typical GDRs is about 10-14 MeV for heavy nuclei~\cite{Berman:1975tt}. Owing to the large GDR cross-section in photonuclear interactions, the exchange of photons between two colliding nuclei in a UPC event has a high probability of exciting one or both ions into GDRs or even higher excitations. GDRs can decay by emitting a single neutron, whereas higher-excitation resonances can decay by emitting two or more neutrons~\cite{Berman:1975tt,STAR:2017enh,Chen:2018tnh}. These neutrons have low momenta, approximately $\sim10$ MeV, comparable to the transverse momentum shift, $\Delta p$, induced by the electromagnetic force. Thus, strong electromagnetic fields may affect the momentum distribution of the emitted neutrons, such that their emission directions correlate with the direction of the impact parameter or the electromagnetic fields. Consequently, a back-to-back correlation may occur between the neutrons emitted from the two colliding nuclei. This provides a means of investigating strong electromagnetic fields in UPCs and offers a way to measure the direction of the impact parameter or electromagnetic field.

\section{Electromagnetic fields}
 The electromagnetic fields can be calculated using the Li\'enard-Wiechert potentials, as described in Refs. ~\cite{Deng:2012pc,Hattori:2016emy}. The electric and magnetic fields are expressed as follows:
\begin{equation}
	\begin{split}
 e{\bm E}(t,\br)=\frac{e^2}{4\pi}\sum_n Z_n\frac{{\bm R}_n-R_n\bv_n}{(R_n-{\bm R}_n\cdot\bv_n)^3}(1-v_n^2),\\
e{\bm B}(t,\br)=\frac{e^2}{4\pi}\sum_n Z_n\frac{\bv_n\times{\bm R}_n}{(R_n-{\bm R}_n\cdot\bv_n)^3}(1-v_n^2),
	\end{split}
	\label{eqEB}
\end{equation}
where $Z_n$ denotes the charge number of the $n$th particle. Here, ${\bm R}_n = \bm r-\bm r_n$ represents the relative coordinates of the field point $\bm r$ to the location $\bm r_n$ of the $n$th particle with velocity $\bm v_n$ at retarded time $t_n = t-|\bm r-\bm r_n|$. Summations were performed for all protons in the nucleus.

For UPCs near the collision time $t = 0$, Eq. (\ref{eqEB}) can be approximated as 
\begin{equation}
	\begin{split}
&e\bE_\perp(0,{\br})\approx Z \frac{e^2}{4\pi}\frac{\sqrt{s_\text{NN}}}{2m_\text{N}}\frac{{\bm R}_{\perp}}
{|{\bm R}_{\perp}|^3},\\
&e\bB_\perp(0,{\br})\approx Z\frac{e^2}{4\pi}\frac{\sqrt{s_\text{NN}}}{2m_\text{N}} \frac{{\bm e}_{z}\times{\bm R}_{\perp}}
{|{\bm R}_{\perp}|^3},
	\end{split}
	\label{eqEB2}
\end{equation}
where ${\bm e}_{z}$ denotes the unit vector in $\pm z$ direction (depending on whether the source nucleus is the target or projectile nucleus), ${\bm R}_{\perp}$ denotes the transverse position of the center of the source nucleus, and $Z$ denotes the total charge number of the source nucleus.

\section{Toy model simulation}

To obtain more quantitative results, MC simulations were performed. We parameterized the nucleon density distribution of the nucleus using the Woods-Saxon function~\cite{Woods:1954zz}:
\begin{equation}
	\rho(r) = \frac{1}{1+{\rm exp}[(r-R)/a]} ,
\label{WSfun}
\end{equation}
where $a=0.54$ fm is the skin depth parameter, and $R = 6.38$ fm is the Au nucleus in this calculation. The electromagnetic fields were calculated using Eq. (\ref{eqEB2})~\cite{Deng:2012pc,Bzdak:2011yy}. We obtained the momentum per nucleon induced by strong electric and Lorentz forces through the following integration:
\begin{equation}
	\begin{split}
		\Delta \bm{p}(b) & = \int^{t=+\infty}_{t=-\infty} \bm{F} \mathrm{d}t  %\\ 
					  % & 
					   = \int^{t=+\infty}_{t=-\infty} (q\bm{E}+ q\bm{v}\times\bm{B}) \mathrm{d}t ,
	\end{split}
\label{momshift}
\end{equation}
where $q = Ze/A$ denotes the average charge per nucleon. Due to the Lorentz contraction, the longitudinal dimension $L_z = 2R/\gamma$ is significantly smaller than the transverse size $2R$, causing the two colliding nuclei to pass each other within a very short timescale $L_z/v_z$. In this study, we assumed that the electromagnetic fields remained constant during the collisions and were negligible before or after the two nuclei touched, a supposition qualitatively consistent with the findings in Ref.\cite{Deng:2012pc}.

Following the methodology of STARlight~\cite{Klein:1999qj,Baltz:2002pp,Baltz:2009jk,Klein:2016yzr}, the probability of an UPC event associated with neutron emission ($P_{xn,xn}^{\rm UPC}$) is calculated as follows:
\begin{equation}
	P_{xn,xn}^{\rm UPC} = P_{0H}(b)\times P_{xn,xn}(b), 
\label{prob0}
\end{equation}
where $P_{0H}(b)$ denotes the probability of having no hadronic interactions and $P_{xn,xn}(b)$ represents the probability of nuclear breakup with neutron emission in both colliding nuclei~\cite{Klein:1999qj,Klein:2016yzr,Broz:2019kpl}. 
Assuming an independent nuclear breakup, $P_{xn,xn}(b)$ can be factorized as
\begin{equation}
		P_{xn,xn}(b)=P_{xn}(b)\times P_{xn}(b).
\label{probBreak0}
\end{equation}
Details of the Eq.\ref{prob0},\ref{probBreak0} can be found in Refs.~\cite{Klein:1999qj,Baltz:2002pp,Baltz:2009jk,Klein:2016yzr} and in the appendix.

Mutual Coulomb dissociation was measured at $\sqrt{s_\text{NN}}$ = 130 GeV Au+Au at the RHIC~\cite{Chiu:2001ij}. The measured cross section of ``Coulomb"-like events agrees well with the theoretical calculations. The measured neutron multiplicity from a ``Coulomb"-like event is concentrated at $\sim 1$, and at most of $\sim 35$. Direct measurement of the neutron multiplicity in UPC events is also available for Au + Au collisions at $\sqrt{s_\text{NN}}$ = 200 GeV from the STAR experiment~\cite{STAR:2007elq,STAR:2017enh}, where the neutron multiplicity distribution is comparable to the above result. 
The Coulomb dissociation of the halo nucleus $^{11}$Be at 72$A$ MeV~\cite{Nakamura:1994zz} shows a cos$(2\phi)$ modulation between the emitted neutron direction and the impact parameter. A cos$(2\phi)$ modulation of the neutrons emitted from the two nuclei in the UPC was proposed to estimate the impact parameter direction~\cite{Baur:2003ar,STAR:ZDCsmd2003}. The current study primarily focuses on the electromagnetic field impact, which mostly manifests as a back-to-back correlation represented by the cos$(\phi)$ modulation. The fluctuation of electromagnetic fields that possess both $x$ and $y$ components may induce higher-order cos$(n\phi)$ modulations. Photoneutron experiments at an energy $\sim$20 MeV measured the polar angle of neutron emission, revealing definite anisotropy on Au and Pb targets~\cite{TAGLIABUE1961144}. Similar results were obtained in other experiments on Coulomb dissociation~\cite{Bakhtiari:2022cbn,Nakamura:1994zz}. Assuming that the photonuclear process is a ``direct photoelectric effect``, an anisotropic polar angular distribution with a maximum at $\pi/2$ for the neutrons is expected~\cite{Courant:1951zz}. Nevertheless, because our study primarily focused on transverse neutron emission, the aforementioned polar angular effect was not considered.

In this study of the neutron emission, the Landau distribution was used to directly estimate the neutron multiplicity from the STAR experiment~\cite{STAR:2007elq,STAR:2017enh}. As mentioned in Ref.~\cite{Broz:2019kpl}, the measurement of spectra for secondary particles from photon-nuclear interactions is currently limited. Therefore, the neutron energies were generated using the same method as in the $\mathrm{n^{O}_{O}n}$ generator~\cite{Broz:2019kpl}, utilizing the Evaluated Nuclear Data File (ENDF) tables~\cite{Chadwick:2011xwu}. ENDF tables for Au are even more limited; tables from $^{208}$Pb are currently used. The accuracy could be improved with future measurements; however, this was sufficient for our qualitative study. The energy distribution of the incident photons in a UPC with neutron emission was calculated as ~\cite{Klein:1999qj,Klein:2016yzr,Broz:2019kpl}
\begin{equation}
\begin{split}
		\frac{\mathrm{d}n(k)}{\mathrm{d}k} &= \int^{+\infty}_{0} 2\pi b db P_{0H}(b) P_{xn,xn}(b) \\
		                         &\times \int^{R}_{0} \frac{r_\perp dr_\perp}{\pi R^{2}} \int^{2\pi}_{0} d\phi \frac{d^{3}N(k,b+r_\perp \cos(\phi))}{dkd^{2}r_\perp}.
		\end{split}
\label{EQphotonK}
\end{equation}

Neutrons are then produced according to the photon energy and ENDF tables with an isotropic angular distribution~\cite{Broz:2019kpl}.

\section{Results and discussions} 
\begin{figure}[hbt]
	\centering
	\includegraphics[width=0.49\textwidth]{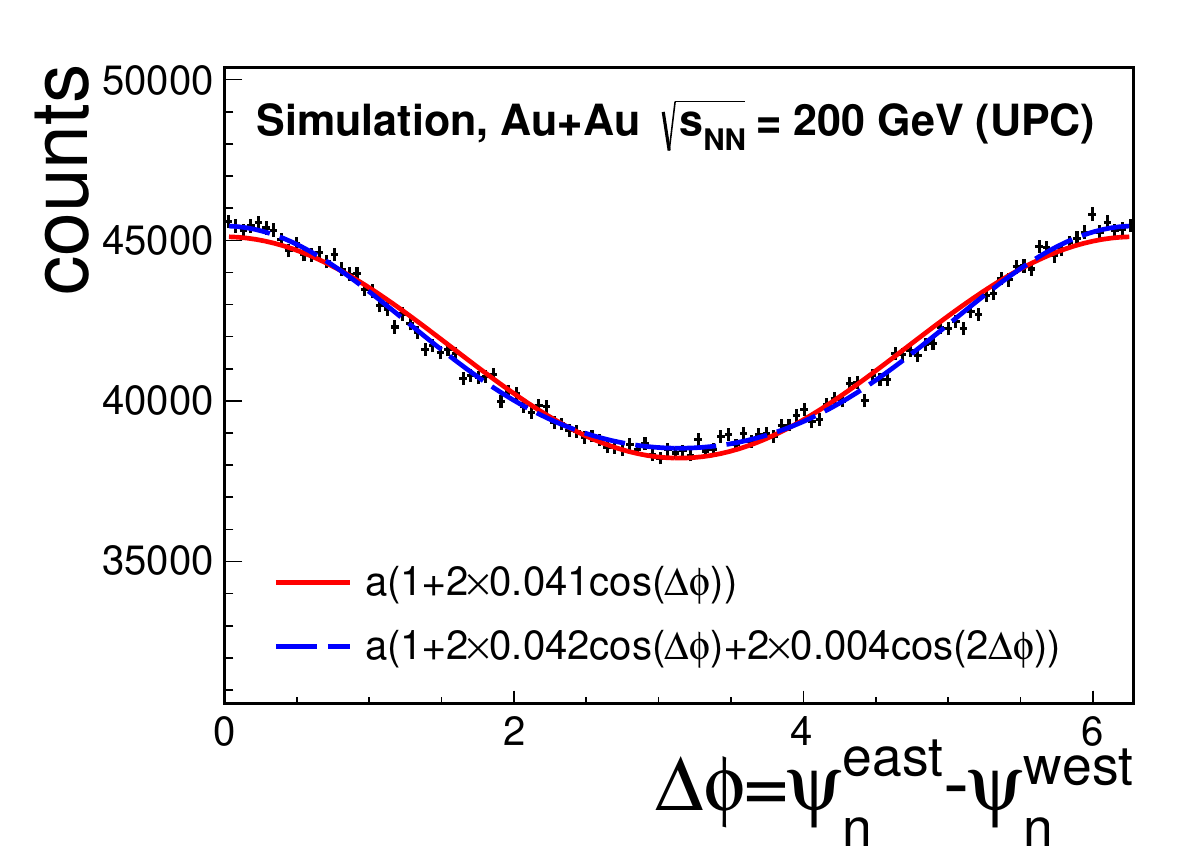}
\caption[]{(Color online) The correlation of the $\Psi^\text{east}_{n}$ and $\Psi^\text{west}_{n}$ defined by the neutrons emitted from the two colliding nuclei (east and west indicate the two nuclei going opposite directions) in $\sqrt{s_\text{NN}}$ = 200 GeV Au + Au UPCs. The $\Psi_n^\text{east}$ is rotated by $\pi$ following experimental conversion. 
	The red line is a cos$(\phi)$ fit to the simulated data, which shows that there are strong electromagnetic forces back-to-back to the two colliding nuclei. 
	Due to the fluctuation, the electromagnetic fields have both $x$ and $y$ components.
	Thus higher-order cos$(2\phi)$ modulation is also included in the fit, as shown by the dashed line.
}
\label{upcEMsimu}
\end{figure}

Figure~\ref{upcEMsimu} presents the simulation results. The $\Psi^\text{east}_{n}$ and $\Psi^\text{west}_{n}$ are the first order harmonic planes~\cite{Poskanzer:1998yz} constructed by the neutrons emitted from the two nuclei (east and west indicate the two nuclei going opposite directions) in $\sqrt{s_{NN}}$ = 200 GeV Au + Au UPCs. 
They are calculated using the sum of the momentum vectors ($q$ vector)~\cite{Poskanzer:1998yz} of neutrons from each nucleus. $\Psi^{east}_{n}$ is rotated by $\pi$ following the experimental conversion. As mentioned in the previous section, the Lorentz force and Coulomb repulsion in such UPCs are sufficiently large to affect the momentum distribution of the emitted neutrons. The red line in Fig.~\ref{upcEMsimu} is a cos$(\phi)$ fit to the simulation, which shows that strong repulsive forces act in opposite directions to the two colliding nuclei. Owing to the fluctuation of electromagnetic fields that possess both $x$ and $y$ components, higher-order cos$(n\phi)$ modulations may also be present. The dashed line represents a fit that includes the cos$(2\phi)$-modulation, which better describes the distribution. Our calculation also demonstrates the sensitivity to the impact parameter direction or the electromagnetic field direction, which could be utilized to study spin-polarization-related physics in UPCs in the future~\cite{Xiao:2020ddm,Xing:2020hwh,Zhang:2020onw,Wu:2022exl}, as the photons are linearly polarized~\cite{STAR:2019wlg}. This may provide insights into the chiral and spin-related effects in relativistic hadronic heavy-ion collisions, such as CME, hyperon spin polarization, and vector meson spin alignment~\cite{Liang:2004ph,STAR:2017ckg,Wang:2023fvy,Chen:2023hnb}.

As discussed in Ref.~\cite{Baur:2003ar,Bertulani:2005ru}, neutrons have been measured using zero-degree calorimeters (ZDCs)~\cite{Adler:2000bd} in relativistic heavy-ion collisions. At the RHIC, the ZDCs are positioned $\pm18$ m away from the nominal interaction point at the center of the detector, with dimensions of 10 cm width and 18.75 cm height~\cite{Adler:2000bd}. For Au + Au collisions at $\sqrt{s_\text{NN}}$ = 200 GeV at the RHIC, where the neutron longitudinal momentum is $\sim$ 100 GeV, a maximum $\pT$ of 140 MeV/c corresponds to a deflection of approximately $2.5$ ~ cm, which is within the acceptance of the ZDCs. With ZDCs' shower-maximum detectors~\cite{STAR:ZDCsmd2003}, which provide position measurements of neutrons, the measurement of the neutron-neutron correlations between the ZDCs is feasible. 

In summary, using Monte Carlo simulation, we demonstrated that the electromagnetic fields generated in UPCs are strong enough to induce measurable back-to-back emission of neutrons in the transverse plane. The effect UPCs discussed here provides a clean way for detecting strong electromagnetic fields, as no hadronic interactions are involved, which would be the strongest electromagnetic fields that can be detected to date. It also offers a method to measure the direction of the impact parameter or electromagnetic field. Thus, it may also shed light on chiral and spin-related effects in relativistic hadronic heavy-ion collisions to understand the fundamental features of hot QCD matter. Additionally, this may aid in understanding the initial conditions of hadronic/heavy-ion collisions with a small $b$, wherein the electromagnetic fields would be considerably stronger. 

%%%%%%%%%%%%%%%%%%%%%%%%%%%%%%%%%%%%%%%%%%%%%%%%%%%%%%%%%%%%%%%%%%%%%%
\section*{Acknowledgments} We thank Dr. Fu-Qiang Wang, Dr. Si-Min Wang, and Dr. Shi Pu for useful discussions.

\section*{Appendix} 

The probability of a UPC event associated with neutron emission ($P_{xn,xn}^{\rm UPC}$) was calculated as follows:
\begin{equation}
	P_{xn,xn}^{\rm UPC} = P_{0H}(b)\times P_{xn,xn}(b), 
\label{prob}
\end{equation}
where $P_{0H}(b)$ denotes the probability of having no hadronic interactions and $P_{xn,xn}(b)$ represents the probability of nuclear breakup with neutron emission in both colliding nuclei~\cite{Klein:1999qj,Klein:2016yzr,Broz:2019kpl}. $P_{0H}(b)$ is given by: 
\begin{equation}
\begin{split}
		&P_{0H}(b) = \exp[-T_{AA}(b)\sigma_{NN}], \\
		&T_{AA}(b) = \int d^{2}\bm{r}_\perp T_{A}(\bm{r}_\perp)T_{A}(\bm{r}_\perp-\bm{b}),
  \end{split}
\label{prob}
\end{equation}
where $\bm b$ is the two-dimensional impact parameter vector (with $b=|\bm b|$) in the transverse plane, $\sigma_{NN}$ is the total nucleon-nucleon interaction cross-section, and $T_{AA}$ is the overlap function. 
The number of nucleon-nucleon collisions follows a Poisson distribution with a mean of $T_{AA}(b)\sigma_{NN}$.
The nuclear thickness function $T_{A}$ is calculated as follows:
\begin{equation}	
		T_{A}(\bm{r}_\perp) = \int dz\; \rho\left(\sqrt{\bm{r}_\perp^{2}+z^{2}}\right),
\label{prob}
\end{equation}
where $\rho$ corresponds to the Woods-Saxon functions in Eq. ~(\ref{WSfun}).

\begin{figure}[hbt]
	\centering
	\includegraphics[width=0.49\textwidth]{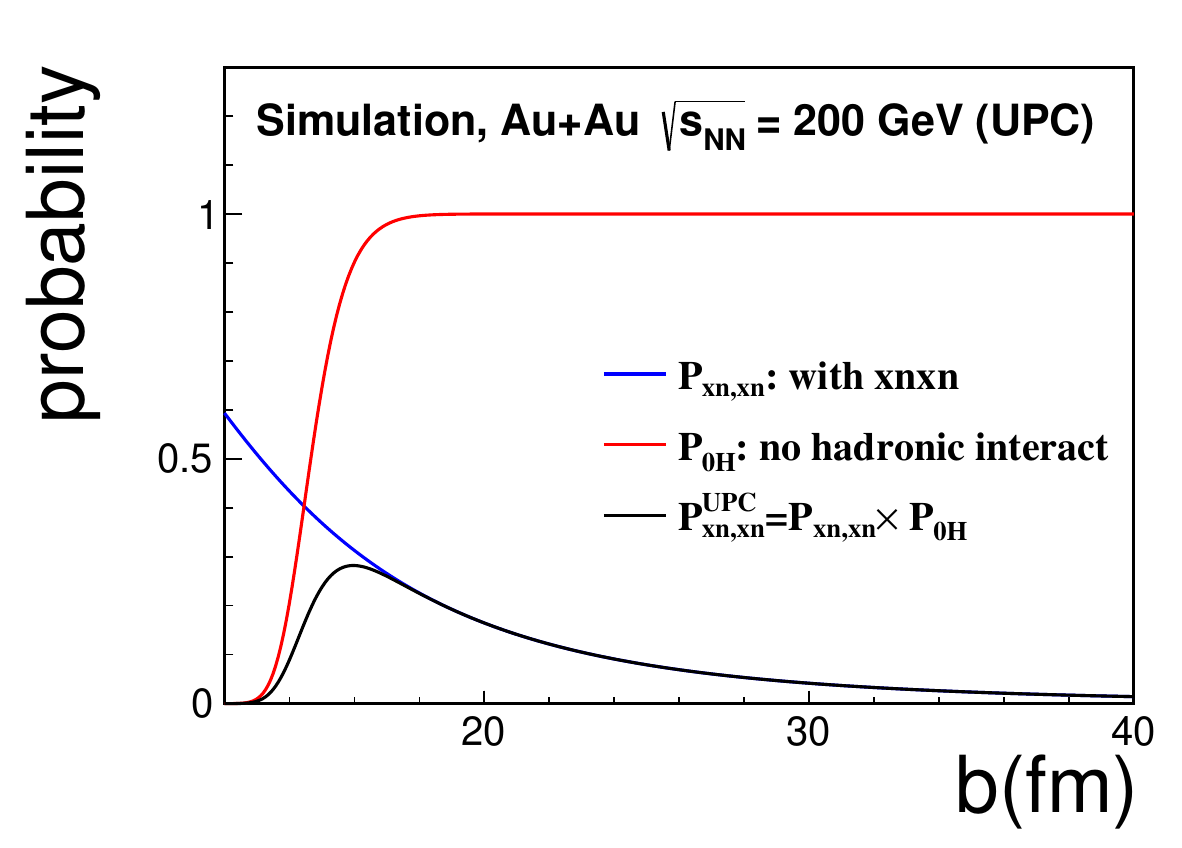}
\caption[]{(Color online) 
The probability of a UPC event associated with neutron emission ($P_{xn,xn}^{\rm UPC}$, black line), the probability of having no hadronic interactions ($P_{0H}(b)$, red line) and the probability of nuclear breakup with neutron emission in both colliding nuclei ($P_{xn,xn}(b)$, blue line) as a function of the impact parameter in Au+Au collisions at $\sqrt{s_\text{NN}}$ = 200 GeV.} 
\label{probability}
\end{figure}

Assuming an independent nuclear breakup, $P_{xn,xn}(b)$ can be factorized as
\begin{equation}
		P_{xn,xn}(b)=P_{xn}(b)\times P_{xn}(b).
\label{probBreak}
\end{equation}
Following the methodology of STARlight~\cite{Klein:1999qj,Baltz:2002pp,Baltz:2009jk,Klein:2016yzr}, the probability of nuclear breakup with neutron emission ($P_{xn}(b)$) is given by
\begin{equation}
		P_{xn}(b)\varpropto \int dk \frac{d^{3}n(b,k)}{dkd^{2}b} \sigma_{\gamma A\rightarrow A^{*}+xn}(k),
\label{probBreak}
\end{equation}
where $\sigma_{\gamma A\rightarrow A^{*}+xn}$ 
was determined from experimental data~\cite{Baltz:1996as,Harland-Lang:2023ohq}. The photon flux was calculated using the Weizs\"acker-Williams approach~\cite{Krauss:1997vr,Klein:1999qj,Klein:2016yzr}
\begin{equation}
		\frac{d^{3}n(r_\perp,k)}{dkd^{2}r_\perp} = \frac{Z^{2}\alpha x^{2}}{{\pi}^{2}kr_\perp^{2}} K_{1}^{2}(x). 
\label{EPA}
\end{equation}
where $k$ represents the photon energy, $Z$ is the nuclear charge, $K_{1}$ is the modified Bessel function, and $x=kr_\perp/\gamma$. Figure ~\ref{probability} shows $P_{0H}(b)$, $P_{xn,xn}(b)$, $P_{xn,xn}^{\rm UPC}$ obtained from the simulation.

\begin{figure}[hbt]
	\centering
	\includegraphics[width=0.49\textwidth]{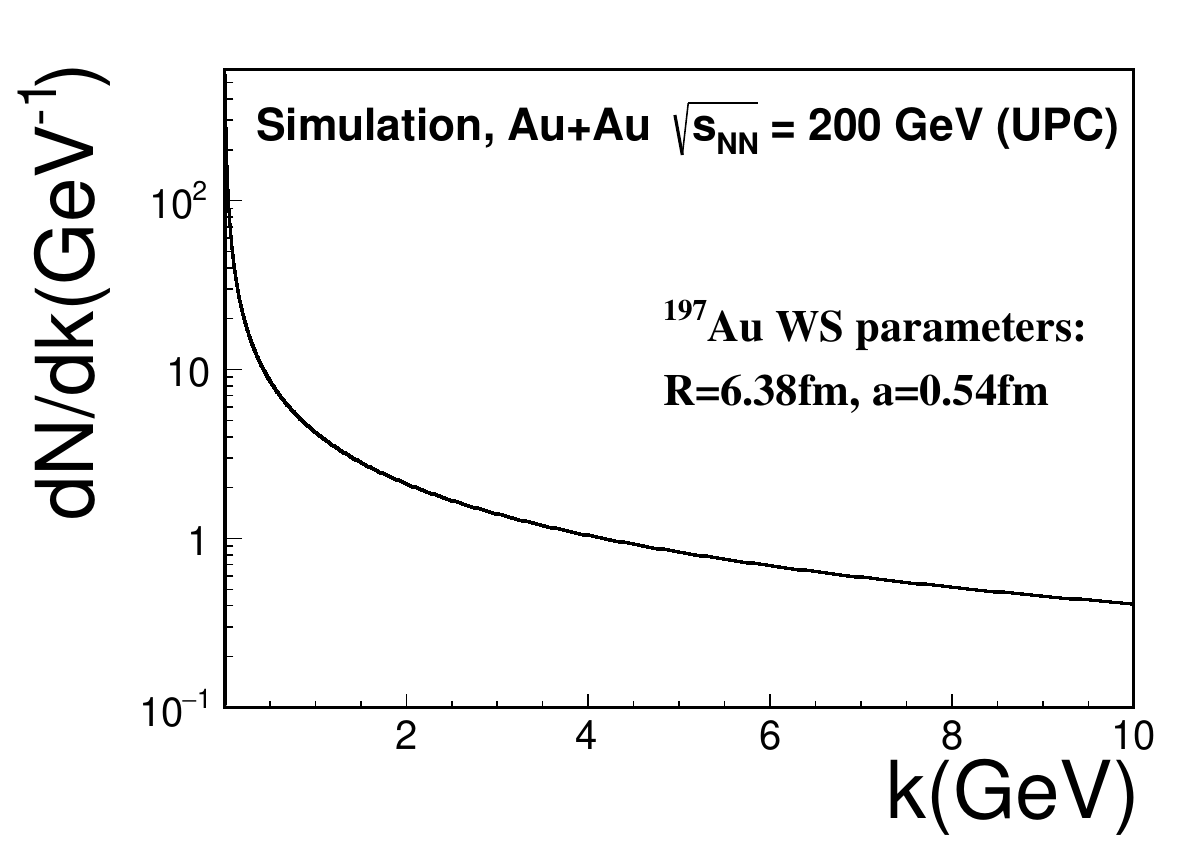}
\caption[]{ Photon energy distribution in the target frame in an UPC event with neutron emission using the Weizs\"acker-Williams approach with Eqs.~(\ref{EPA}) and (\ref{EQphotonK}).
}
\label{photonK}
\end{figure}

Figure~\ref{photonK} shows the calculation of the photon energy distribution in the target frame for a UPC event with neutron emissions.

\bibliographystyle{nst}
\bibliography{ref}

\begin{thebibliography}{58}%
\makeatletter
\providecommand \@ifxundefined [1]{%
 \@ifx{#1\undefined}
}%
\providecommand \@ifnum [1]{%
 \ifnum #1\expandafter \@firstoftwo
 \else \expandafter \@secondoftwo
 \fi
}%
\providecommand \@ifx [1]{%
 \ifx #1\expandafter \@firstoftwo
 \else \expandafter \@secondoftwo
 \fi
}%
\providecommand \natexlab [1]{#1}%
\providecommand \enquote  [1]{``#1''}%
\providecommand \bibnamefont  [1]{#1}%
\providecommand \bibfnamefont [1]{#1}%
\providecommand \citenamefont [1]{#1}%
\providecommand \href@noop [0]{\@secondoftwo}%
\providecommand \href [0]{\begingroup \@sanitize@url \@href}%
\providecommand \@href[1]{\@@startlink{#1}\@@href}%
\providecommand \@@href[1]{\endgroup#1\@@endlink}%
\providecommand \@sanitize@url [0]{\catcode `\\12\catcode `\$12\catcode
  `\&12\catcode `\#12\catcode `\^12\catcode `\_12\catcode `\%12\relax}%
\providecommand \@@startlink[1]{}%
\providecommand \@@endlink[0]{}%
\providecommand \url  [0]{\begingroup\@sanitize@url \@url }%
\providecommand \@url [1]{\endgroup\@href {#1}{\urlprefix }}%
\providecommand \urlprefix  [0]{URL }%
\providecommand \Eprint [0]{\href }%
\providecommand \doibase [0]{http://dx.doi.org/}%
\providecommand \selectlanguage [0]{\@gobble}%
\providecommand \bibinfo  [0]{\@secondoftwo}%
\providecommand \bibfield  [0]{\@secondoftwo}%
\providecommand \translation [1]{[#1]}%
\providecommand \BibitemOpen [0]{}%
\providecommand \bibitemStop [0]{}%
\providecommand \bibitemNoStop [0]{.\EOS\space}%
\providecommand \EOS [0]{\spacefactor3000\relax}%
\providecommand \BibitemShut  [1]{\csname bibitem#1\endcsname}%
\let\auto@bib@innerbib\@empty
%</preamble>
\bibitem [{\citenamefont {Kharzeev}\ \emph {et~al.}(1998)\citenamefont
  {Kharzeev}, \citenamefont {Pisarski},\ and\ \citenamefont
  {Tytgat}}]{Kharzeev:1998kz}%
  \BibitemOpen
  \bibfield  {author} {\bibinfo {author} {\bibfnamefont {D.}~\bibnamefont
  {Kharzeev}}, \bibinfo {author} {\bibfnamefont {R.~D.}\ \bibnamefont
  {Pisarski}}, \ and\ \bibinfo {author} {\bibfnamefont {M.~H.~G.}\ \bibnamefont
  {Tytgat}},\ }\href {\doibase https://doi.org/10.1103/PhysRevLett.81.512}
  {\bibfield  {journal} {\bibinfo  {journal} {Phys. Rev. Lett.}\ }\textbf
  {\bibinfo {volume} {81}},\ \bibinfo {pages} {512} (\bibinfo {year}
  {1998})}\BibitemShut {NoStop}%
\bibitem [{\citenamefont {Huang}(2016)}]{Huang:2015oca}%
  \BibitemOpen
  \bibfield  {author} {\bibinfo {author} {\bibfnamefont {X.-G.}\ \bibnamefont
  {Huang}},\ }\href {\doibase 10.1088/0034-4885/79/7/076302} {\bibfield
  {journal} {\bibinfo  {journal} {Rept. Prog. Phys.}\ }\textbf {\bibinfo
  {volume} {79}},\ \bibinfo {pages} {076302} (\bibinfo {year}
  {2016})}\BibitemShut {NoStop}%
\bibitem [{\citenamefont {Wang}\ and\ \citenamefont
  {Zhao}(2018)}]{Wang:2018ygc}%
  \BibitemOpen
  \bibfield  {author} {\bibinfo {author} {\bibfnamefont {F.-Q.}\ \bibnamefont
  {Wang}}\ and\ \bibinfo {author} {\bibfnamefont {J.}~\bibnamefont {Zhao}},\
  }\href {\doibase 10.1007/s41365-018-0520-z} {\bibfield  {journal} {\bibinfo
  {journal} {Nucl. Sci. Tech.}\ }\textbf {\bibinfo {volume} {29}},\ \bibinfo
  {pages} {179} (\bibinfo {year} {2018})}\BibitemShut {NoStop}%
\bibitem [{\citenamefont {Zhao}\ and\ \citenamefont
  {Wang}(2019)}]{Zhao:2019hta}%
  \BibitemOpen
  \bibfield  {author} {\bibinfo {author} {\bibfnamefont {J.}~\bibnamefont
  {Zhao}}\ and\ \bibinfo {author} {\bibfnamefont {F.}~\bibnamefont {Wang}},\
  }\href {\doibase 10.1016/j.ppnp.2019.05.001} {\bibfield  {journal} {\bibinfo
  {journal} {Prog. Part. Nucl. Phys.}\ }\textbf {\bibinfo {volume} {107}},\
  \bibinfo {pages} {200} (\bibinfo {year} {2019})}\BibitemShut {NoStop}%
\bibitem [{\citenamefont {Fang}\ \emph {et~al.}(2021)\citenamefont {Fang},
  \citenamefont {Dong}, \citenamefont {Hou},\ and\ \citenamefont
  {Sun}}]{Fang:2021cpl}%
  \BibitemOpen
  \bibfield  {author} {\bibinfo {author} {\bibfnamefont {R.}~\bibnamefont
  {Fang}}, \bibinfo {author} {\bibfnamefont {R.}~\bibnamefont {Dong}}, \bibinfo
  {author} {\bibfnamefont {D.}~\bibnamefont {Hou}}, \ and\ \bibinfo {author}
  {\bibfnamefont {B.}~\bibnamefont {Sun}},\ }\href {\doibase
  10.1088/0256-307X/38/9/091201} {\bibfield  {journal} {\bibinfo  {journal}
  {Chin. Phys. Lett.}\ }\textbf {\bibinfo {volume} {38}},\ \bibinfo {pages}
  {091201} (\bibinfo {year} {2021})}\BibitemShut {NoStop}%
\bibitem [{\citenamefont {Liu}\ and\ \citenamefont
  {Huang}(2022)}]{Liu:2022Sci}%
  \BibitemOpen
  \bibfield  {author} {\bibinfo {author} {\bibfnamefont {Y.-C.}\ \bibnamefont
  {Liu}}\ and\ \bibinfo {author} {\bibfnamefont {X.-G.}\ \bibnamefont
  {Huang}},\ }\href {\doibase 10.1007/s11433-022-1903-8} {\bibfield  {journal}
  {\bibinfo  {journal} {SCIENCE CHINA-PHYSICS MECHANICS and ASTRONOMY}\
  }\textbf {\bibinfo {volume} {65}},\ \bibinfo {pages} {272011} (\bibinfo
  {year} {2022})}\BibitemShut {NoStop}%
\bibitem [{\citenamefont {Adamczyk}\ \emph
  {et~al.}(2017{\natexlab{a}})\citenamefont {Adamczyk} \emph
  {et~al.}}]{STAR:2017ckg}%
  \BibitemOpen
  \bibfield  {author} {\bibinfo {author} {\bibfnamefont {L.}~\bibnamefont
  {Adamczyk}} \emph {et~al.} (\bibinfo {collaboration} {STAR}),\ }\href
  {\doibase 10.1038/nature23004} {\bibfield  {journal} {\bibinfo  {journal}
  {Nature}\ }\textbf {\bibinfo {volume} {548}},\ \bibinfo {pages} {62}
  (\bibinfo {year} {2017}{\natexlab{a}})},\ \Eprint
  {http://arxiv.org/abs/1701.06657} {arXiv:1701.06657 [nucl-ex]} \BibitemShut
  {NoStop}%
\bibitem [{\citenamefont {Adam}\ \emph
  {et~al.}(2021{\natexlab{a}})\citenamefont {Adam}, \citenamefont {Adamczyk},
  \citenamefont {Zyzak} \emph {et~al.}}]{STAR:2021nst}%
  \BibitemOpen
  \bibfield  {author} {\bibinfo {author} {\bibfnamefont {J.}~\bibnamefont
  {Adam}}, \bibinfo {author} {\bibfnamefont {L.}~\bibnamefont {Adamczyk}},
  \bibinfo {author} {\bibfnamefont {M.}~\bibnamefont {Zyzak}},  \emph
  {et~al.},\ }\href {\doibase 10.1007/s41365-021-00878-y} {\bibfield  {journal}
  {\bibinfo  {journal} {Nucl. Sci. Tech.}\ }\textbf {\bibinfo {volume} {32}},\
  \bibinfo {pages} {48} (\bibinfo {year} {2021}{\natexlab{a}})}\BibitemShut
  {NoStop}%
\bibitem [{\citenamefont {Liu}\ and\ \citenamefont
  {Huang}(2020)}]{Liu:2020nst}%
  \BibitemOpen
  \bibfield  {author} {\bibinfo {author} {\bibfnamefont {Y.~C.}\ \bibnamefont
  {Liu}}\ and\ \bibinfo {author} {\bibfnamefont {X.~G.}\ \bibnamefont
  {Huang}},\ }\href {\doibase 10.1007/s41365-020-00764-z} {\bibfield  {journal}
  {\bibinfo  {journal} {Nucl. Sci. Tech.}\ }\textbf {\bibinfo {volume} {31}},\
  \bibinfo {pages} {56} (\bibinfo {year} {2020})}\BibitemShut {NoStop}%
\bibitem [{\citenamefont {Gao}\ \emph {et~al.}(2020)\citenamefont {Gao},
  \citenamefont {Ma}, \citenamefont {Pu},\ and\ \citenamefont
  {Wang}}]{Gao:2020nst}%
  \BibitemOpen
  \bibfield  {author} {\bibinfo {author} {\bibfnamefont {J.-H.}\ \bibnamefont
  {Gao}}, \bibinfo {author} {\bibfnamefont {G.-L.}\ \bibnamefont {Ma}},
  \bibinfo {author} {\bibfnamefont {S.}~\bibnamefont {Pu}}, \ and\ \bibinfo
  {author} {\bibfnamefont {Q.}~\bibnamefont {Wang}},\ }\href {\doibase
  10.1007/s41365-020-00801-x} {\bibfield  {journal} {\bibinfo  {journal} {Nucl.
  Sci. Tech.}\ }\textbf {\bibinfo {volume} {31}},\ \bibinfo {pages} {9}
  (\bibinfo {year} {2020})}\BibitemShut {NoStop}%
\bibitem [{\citenamefont {Bertulani}\ \emph {et~al.}(2005)\citenamefont
  {Bertulani}, \citenamefont {Klein},\ and\ \citenamefont
  {Nystrand}}]{Bertulani:2005ru}%
  \BibitemOpen
  \bibfield  {author} {\bibinfo {author} {\bibfnamefont {C.~A.}\ \bibnamefont
  {Bertulani}}, \bibinfo {author} {\bibfnamefont {S.~R.}\ \bibnamefont
  {Klein}}, \ and\ \bibinfo {author} {\bibfnamefont {J.}~\bibnamefont
  {Nystrand}},\ }\href {\doibase 10.1146/annurev.nucl.55.090704.151526}
  {\bibfield  {journal} {\bibinfo  {journal} {Ann. Rev. Nucl. Part. Sci.}\
  }\textbf {\bibinfo {volume} {55}},\ \bibinfo {pages} {271} (\bibinfo {year}
  {2005})},\ \Eprint {http://arxiv.org/abs/nucl-ex/0502005}
  {arXiv:nucl-ex/0502005} \BibitemShut {NoStop}%
\bibitem [{\citenamefont {Klein}\ \emph {et~al.}(2017)\citenamefont {Klein},
  \citenamefont {Nystrand}, \citenamefont {Seger}, \citenamefont {Gorbunov},\
  and\ \citenamefont {Butterworth}}]{Klein:2016yzr}%
  \BibitemOpen
  \bibfield  {author} {\bibinfo {author} {\bibfnamefont {S.~R.}\ \bibnamefont
  {Klein}}, \bibinfo {author} {\bibfnamefont {J.}~\bibnamefont {Nystrand}},
  \bibinfo {author} {\bibfnamefont {J.}~\bibnamefont {Seger}}, \bibinfo
  {author} {\bibfnamefont {Y.}~\bibnamefont {Gorbunov}}, \ and\ \bibinfo
  {author} {\bibfnamefont {J.}~\bibnamefont {Butterworth}},\ }\href {\doibase
  10.1016/j.cpc.2016.10.016} {\bibfield  {journal} {\bibinfo  {journal}
  {Comput. Phys. Commun.}\ }\textbf {\bibinfo {volume} {212}},\ \bibinfo
  {pages} {258} (\bibinfo {year} {2017})},\ \Eprint
  {http://arxiv.org/abs/1607.03838} {arXiv:1607.03838 [hep-ph]} \BibitemShut
  {NoStop}%
\bibitem [{\citenamefont {Klein}\ and\ \citenamefont
  {Steinberg}(2020)}]{Klein:2020fmr}%
  \BibitemOpen
  \bibfield  {author} {\bibinfo {author} {\bibfnamefont {S.}~\bibnamefont
  {Klein}}\ and\ \bibinfo {author} {\bibfnamefont {P.}~\bibnamefont
  {Steinberg}},\ }\href {\doibase 10.1146/annurev-nucl-030320-033923}
  {\bibfield  {journal} {\bibinfo  {journal} {Ann. Rev. Nucl. Part. Sci.}\
  }\textbf {\bibinfo {volume} {70}},\ \bibinfo {pages} {323} (\bibinfo {year}
  {2020})},\ \Eprint {http://arxiv.org/abs/2005.01872} {arXiv:2005.01872
  [nucl-ex]} \BibitemShut {NoStop}%
\bibitem [{\citenamefont {Baur}\ \emph {et~al.}(1998)\citenamefont {Baur},
  \citenamefont {Hencken},\ and\ \citenamefont {Trautmann}}]{Baur:1998ay}%
  \BibitemOpen
  \bibfield  {author} {\bibinfo {author} {\bibfnamefont {G.}~\bibnamefont
  {Baur}}, \bibinfo {author} {\bibfnamefont {K.}~\bibnamefont {Hencken}}, \
  and\ \bibinfo {author} {\bibfnamefont {D.}~\bibnamefont {Trautmann}},\ }\href
  {\doibase 10.1088/0954-3899/24/9/003} {\bibfield  {journal} {\bibinfo
  {journal} {J. Phys. G}\ }\textbf {\bibinfo {volume} {24}},\ \bibinfo {pages}
  {1657} (\bibinfo {year} {1998})},\ \Eprint
  {http://arxiv.org/abs/hep-ph/9804348} {arXiv:hep-ph/9804348} \BibitemShut
  {NoStop}%
\bibitem [{\citenamefont {Klein}\ and\ \citenamefont
  {Nystrand}(1999)}]{Klein:1999qj}%
  \BibitemOpen
  \bibfield  {author} {\bibinfo {author} {\bibfnamefont {S.}~\bibnamefont
  {Klein}}\ and\ \bibinfo {author} {\bibfnamefont {J.}~\bibnamefont
  {Nystrand}},\ }\href {\doibase 10.1103/PhysRevC.60.014903} {\bibfield
  {journal} {\bibinfo  {journal} {Phys. Rev. C}\ }\textbf {\bibinfo {volume}
  {60}},\ \bibinfo {pages} {014903} (\bibinfo {year} {1999})},\ \Eprint
  {http://arxiv.org/abs/hep-ph/9902259} {arXiv:hep-ph/9902259} \BibitemShut
  {NoStop}%
\bibitem [{\citenamefont {Adler}\ \emph {et~al.}(2002)\citenamefont {Adler}
  \emph {et~al.}}]{STAR:2002caw}%
  \BibitemOpen
  \bibfield  {author} {\bibinfo {author} {\bibfnamefont {C.}~\bibnamefont
  {Adler}} \emph {et~al.} (\bibinfo {collaboration} {STAR}),\ }\href {\doibase
  10.1103/PhysRevLett.89.272302} {\bibfield  {journal} {\bibinfo  {journal}
  {Phys. Rev. Lett.}\ }\textbf {\bibinfo {volume} {89}},\ \bibinfo {pages}
  {272302} (\bibinfo {year} {2002})},\ \Eprint
  {http://arxiv.org/abs/nucl-ex/0206004} {arXiv:nucl-ex/0206004} \BibitemShut
  {NoStop}%
\bibitem [{\citenamefont {Afanasiev}\ \emph {et~al.}(2009)\citenamefont
  {Afanasiev} \emph {et~al.}}]{PHENIX:2009xtn}%
  \BibitemOpen
  \bibfield  {author} {\bibinfo {author} {\bibfnamefont {S.}~\bibnamefont
  {Afanasiev}} \emph {et~al.} (\bibinfo {collaboration} {PHENIX}),\ }\href
  {\doibase 10.1016/j.physletb.2009.07.061} {\bibfield  {journal} {\bibinfo
  {journal} {Phys. Lett. B}\ }\textbf {\bibinfo {volume} {679}},\ \bibinfo
  {pages} {321} (\bibinfo {year} {2009})},\ \Eprint
  {http://arxiv.org/abs/0903.2041} {arXiv:0903.2041 [nucl-ex]} \BibitemShut
  {NoStop}%
\bibitem [{\citenamefont {Abelev}\ \emph {et~al.}(2013)\citenamefont {Abelev}
  \emph {et~al.}}]{ALICE:2012yye}%
  \BibitemOpen
  \bibfield  {author} {\bibinfo {author} {\bibfnamefont {B.}~\bibnamefont
  {Abelev}} \emph {et~al.} (\bibinfo {collaboration} {ALICE}),\ }\href
  {\doibase 10.1016/j.physletb.2012.11.059} {\bibfield  {journal} {\bibinfo
  {journal} {Phys. Lett. B}\ }\textbf {\bibinfo {volume} {718}},\ \bibinfo
  {pages} {1273} (\bibinfo {year} {2013})},\ \Eprint
  {http://arxiv.org/abs/1209.3715} {arXiv:1209.3715 [nucl-ex]} \BibitemShut
  {NoStop}%
\bibitem [{\citenamefont {Khachatryan}\ \emph {et~al.}(2017)\citenamefont
  {Khachatryan} \emph {et~al.}}]{CMS:2016itn}%
  \BibitemOpen
  \bibfield  {author} {\bibinfo {author} {\bibfnamefont {V.}~\bibnamefont
  {Khachatryan}} \emph {et~al.} (\bibinfo {collaboration} {CMS}),\ }\href
  {\doibase 10.1016/j.physletb.2017.07.001} {\bibfield  {journal} {\bibinfo
  {journal} {Phys. Lett. B}\ }\textbf {\bibinfo {volume} {772}},\ \bibinfo
  {pages} {489} (\bibinfo {year} {2017})},\ \Eprint
  {http://arxiv.org/abs/1605.06966} {arXiv:1605.06966 [nucl-ex]} \BibitemShut
  {NoStop}%
\bibitem [{\citenamefont {Aaboud}\ \emph {et~al.}(2017)\citenamefont {Aaboud}
  \emph {et~al.}}]{ATLAS:2017fur}%
  \BibitemOpen
  \bibfield  {author} {\bibinfo {author} {\bibfnamefont {M.}~\bibnamefont
  {Aaboud}} \emph {et~al.} (\bibinfo {collaboration} {ATLAS}),\ }\href
  {\doibase 10.1038/nphys4208} {\bibfield  {journal} {\bibinfo  {journal}
  {Nature Phys.}\ }\textbf {\bibinfo {volume} {13}},\ \bibinfo {pages} {852}
  (\bibinfo {year} {2017})},\ \Eprint {http://arxiv.org/abs/1702.01625}
  {arXiv:1702.01625 [hep-ex]} \BibitemShut {NoStop}%
\bibitem [{\citenamefont {Adam}\ \emph
  {et~al.}(2021{\natexlab{b}})\citenamefont {Adam} \emph
  {et~al.}}]{STAR:2019wlg}%
  \BibitemOpen
  \bibfield  {author} {\bibinfo {author} {\bibfnamefont {J.}~\bibnamefont
  {Adam}} \emph {et~al.} (\bibinfo {collaboration} {STAR}),\ }\href {\doibase
  10.1103/PhysRevLett.127.052302} {\bibfield  {journal} {\bibinfo  {journal}
  {Phys. Rev. Lett.}\ }\textbf {\bibinfo {volume} {127}},\ \bibinfo {pages}
  {052302} (\bibinfo {year} {2021}{\natexlab{b}})},\ \Eprint
  {http://arxiv.org/abs/1910.12400} {arXiv:1910.12400 [nucl-ex]} \BibitemShut
  {NoStop}%
\bibitem [{\citenamefont {Goldhaber}\ and\ \citenamefont
  {Teller}(1948)}]{Goldhaber:1948zza}%
  \BibitemOpen
  \bibfield  {author} {\bibinfo {author} {\bibfnamefont {M.}~\bibnamefont
  {Goldhaber}}\ and\ \bibinfo {author} {\bibfnamefont {E.}~\bibnamefont
  {Teller}},\ }\href {\doibase 10.1103/PhysRev.74.1046} {\bibfield  {journal}
  {\bibinfo  {journal} {Phys. Rev.}\ }\textbf {\bibinfo {volume} {74}},\
  \bibinfo {pages} {1046} (\bibinfo {year} {1948})}\BibitemShut {NoStop}%
\bibitem [{\citenamefont {Berman}\ and\ \citenamefont
  {Fultz}(1975)}]{Berman:1975tt}%
  \BibitemOpen
  \bibfield  {author} {\bibinfo {author} {\bibfnamefont {B.~L.}\ \bibnamefont
  {Berman}}\ and\ \bibinfo {author} {\bibfnamefont {S.~C.}\ \bibnamefont
  {Fultz}},\ }\href {\doibase 10.1103/RevModPhys.47.713} {\bibfield  {journal}
  {\bibinfo  {journal} {Rev. Mod. Phys.}\ }\textbf {\bibinfo {volume} {47}},\
  \bibinfo {pages} {713} (\bibinfo {year} {1975})}\BibitemShut {NoStop}%
\bibitem [{\citenamefont {Chomaz}\ and\ \citenamefont
  {Frascaria}(1995)}]{Chomaz:1993qe}%
  \BibitemOpen
  \bibfield  {author} {\bibinfo {author} {\bibfnamefont {P.}~\bibnamefont
  {Chomaz}}\ and\ \bibinfo {author} {\bibfnamefont {N.}~\bibnamefont
  {Frascaria}},\ }\href {\doibase 10.1016/0370-1573(94)00079-I} {\bibfield
  {journal} {\bibinfo  {journal} {Phys. Rept.}\ }\textbf {\bibinfo {volume}
  {252}},\ \bibinfo {pages} {275} (\bibinfo {year} {1995})}\BibitemShut
  {NoStop}%
\bibitem [{\citenamefont {Veyssiere}\ \emph {et~al.}(1970)\citenamefont
  {Veyssiere}, \citenamefont {Beil}, \citenamefont {Bergere}, \citenamefont
  {Carlos},\ and\ \citenamefont {Lepretre}}]{Veyssiere:1970ztg}%
  \BibitemOpen
  \bibfield  {author} {\bibinfo {author} {\bibfnamefont {A.}~\bibnamefont
  {Veyssiere}}, \bibinfo {author} {\bibfnamefont {H.}~\bibnamefont {Beil}},
  \bibinfo {author} {\bibfnamefont {R.}~\bibnamefont {Bergere}}, \bibinfo
  {author} {\bibfnamefont {P.}~\bibnamefont {Carlos}}, \ and\ \bibinfo {author}
  {\bibfnamefont {A.}~\bibnamefont {Lepretre}},\ }\href {\doibase
  10.1016/0375-9474(70)90727-X} {\bibfield  {journal} {\bibinfo  {journal}
  {Nucl. Phys. A}\ }\textbf {\bibinfo {volume} {159}},\ \bibinfo {pages} {561}
  (\bibinfo {year} {1970})}\BibitemShut {NoStop}%
\bibitem [{\citenamefont {Tao}\ \emph {et~al.}(2013)\citenamefont {Tao},
  \citenamefont {Ma}, \citenamefont {Zhang}, \citenamefont {Cao}, \citenamefont
  {Fang},\ and\ \citenamefont {Wang}}]{Tao:2013PRC}%
  \BibitemOpen
  \bibfield  {author} {\bibinfo {author} {\bibfnamefont {C.}~\bibnamefont
  {Tao}}, \bibinfo {author} {\bibfnamefont {Y.~G.}\ \bibnamefont {Ma}},
  \bibinfo {author} {\bibfnamefont {G.~Q.}\ \bibnamefont {Zhang}}, \bibinfo
  {author} {\bibfnamefont {X.~G.}\ \bibnamefont {Cao}}, \bibinfo {author}
  {\bibfnamefont {D.~Q.}\ \bibnamefont {Fang}}, \ and\ \bibinfo {author}
  {\bibfnamefont {H.~W.}\ \bibnamefont {Wang}},\ }\href {\doibase
  10.1103/PhysRevC.87.014621} {\bibfield  {journal} {\bibinfo  {journal} {Phys.
  Rev. C}\ }\textbf {\bibinfo {volume} {87}},\ \bibinfo {pages} {014621}
  (\bibinfo {year} {2013})}\BibitemShut {NoStop}%
\bibitem [{\citenamefont {He}\ \emph {et~al.}(2014)\citenamefont {He},
  \citenamefont {Ma}, \citenamefont {Cao}, \citenamefont {Cai},\ and\
  \citenamefont {Zhang}}]{He:2014PRL}%
  \BibitemOpen
  \bibfield  {author} {\bibinfo {author} {\bibfnamefont {W.~B.}\ \bibnamefont
  {He}}, \bibinfo {author} {\bibfnamefont {Y.~G.}\ \bibnamefont {Ma}}, \bibinfo
  {author} {\bibfnamefont {X.~G.}\ \bibnamefont {Cao}}, \bibinfo {author}
  {\bibfnamefont {X.~Z.}\ \bibnamefont {Cai}}, \ and\ \bibinfo {author}
  {\bibfnamefont {G.~Q.}\ \bibnamefont {Zhang}},\ }\href {\doibase
  10.1103/PhysRevLett.113.032506} {\bibfield  {journal} {\bibinfo  {journal}
  {Phys. Rev. Lett.}\ }\textbf {\bibinfo {volume} {113}},\ \bibinfo {pages}
  {032506} (\bibinfo {year} {2014})}\BibitemShut {NoStop}%
\bibitem [{\citenamefont {Huang}\ and\ \citenamefont
  {Ma}(2021)}]{Huang:2021PRC}%
  \BibitemOpen
  \bibfield  {author} {\bibinfo {author} {\bibfnamefont {B.-S.}\ \bibnamefont
  {Huang}}\ and\ \bibinfo {author} {\bibfnamefont {Y.-G.}\ \bibnamefont {Ma}},\
  }\href {\doibase 10.1103/PhysRevC.103.054318} {\bibfield  {journal} {\bibinfo
   {journal} {Phys. Rev. C}\ }\textbf {\bibinfo {volume} {103}},\ \bibinfo
  {pages} {054318} (\bibinfo {year} {2021})}\BibitemShut {NoStop}%
\bibitem [{\citenamefont {Baltz}\ \emph {et~al.}(2002)\citenamefont {Baltz},
  \citenamefont {Klein},\ and\ \citenamefont {Nystrand}}]{Baltz:2002pp}%
  \BibitemOpen
  \bibfield  {author} {\bibinfo {author} {\bibfnamefont {A.~J.}\ \bibnamefont
  {Baltz}}, \bibinfo {author} {\bibfnamefont {S.~R.}\ \bibnamefont {Klein}}, \
  and\ \bibinfo {author} {\bibfnamefont {J.}~\bibnamefont {Nystrand}},\ }\href
  {\doibase 10.1103/PhysRevLett.89.012301} {\bibfield  {journal} {\bibinfo
  {journal} {Phys. Rev. Lett.}\ }\textbf {\bibinfo {volume} {89}},\ \bibinfo
  {pages} {012301} (\bibinfo {year} {2002})},\ \Eprint
  {http://arxiv.org/abs/nucl-th/0205031} {arXiv:nucl-th/0205031} \BibitemShut
  {NoStop}%
\bibitem [{\citenamefont {Deng}\ and\ \citenamefont
  {Huang}(2012)}]{Deng:2012pc}%
  \BibitemOpen
  \bibfield  {author} {\bibinfo {author} {\bibfnamefont {W.-T.}\ \bibnamefont
  {Deng}}\ and\ \bibinfo {author} {\bibfnamefont {X.-G.}\ \bibnamefont
  {Huang}},\ }\href {\doibase 10.1103/PhysRevC.85.044907} {\bibfield  {journal}
  {\bibinfo  {journal} {Phys. Rev. C}\ }\textbf {\bibinfo {volume} {85}},\
  \bibinfo {pages} {044907} (\bibinfo {year} {2012})},\ \Eprint
  {http://arxiv.org/abs/1201.5108} {arXiv:1201.5108 [nucl-th]} \BibitemShut
  {NoStop}%
\bibitem [{\citenamefont {Bzdak}\ and\ \citenamefont
  {Skokov}(2012)}]{Bzdak:2011yy}%
  \BibitemOpen
  \bibfield  {author} {\bibinfo {author} {\bibfnamefont {A.}~\bibnamefont
  {Bzdak}}\ and\ \bibinfo {author} {\bibfnamefont {V.}~\bibnamefont {Skokov}},\
  }\href {\doibase 10.1016/j.physletb.2012.02.065} {\bibfield  {journal}
  {\bibinfo  {journal} {Phys. Lett. B}\ }\textbf {\bibinfo {volume} {710}},\
  \bibinfo {pages} {171} (\bibinfo {year} {2012})},\ \Eprint
  {http://arxiv.org/abs/1111.1949} {arXiv:1111.1949 [hep-ph]} \BibitemShut
  {NoStop}%
\bibitem [{\citenamefont {Adamczyk}\ \emph
  {et~al.}(2017{\natexlab{b}})\citenamefont {Adamczyk} \emph
  {et~al.}}]{STAR:2017enh}%
  \BibitemOpen
  \bibfield  {author} {\bibinfo {author} {\bibfnamefont {L.}~\bibnamefont
  {Adamczyk}} \emph {et~al.} (\bibinfo {collaboration} {STAR}),\ }\href
  {\doibase 10.1103/PhysRevC.96.054904} {\bibfield  {journal} {\bibinfo
  {journal} {Phys. Rev. C}\ }\textbf {\bibinfo {volume} {96}},\ \bibinfo
  {pages} {054904} (\bibinfo {year} {2017}{\natexlab{b}})},\ \Eprint
  {http://arxiv.org/abs/1702.07705} {arXiv:1702.07705 [nucl-ex]} \BibitemShut
  {NoStop}%
\bibitem [{\citenamefont {Chen}\ \emph {et~al.}(2018)\citenamefont {Chen},
  \citenamefont {Keane}, \citenamefont {Ma}, \citenamefont {Tang},\ and\
  \citenamefont {Xu}}]{Chen:2018tnh}%
  \BibitemOpen
  \bibfield  {author} {\bibinfo {author} {\bibfnamefont {J.}~\bibnamefont
  {Chen}}, \bibinfo {author} {\bibfnamefont {D.}~\bibnamefont {Keane}},
  \bibinfo {author} {\bibfnamefont {Y.-G.}\ \bibnamefont {Ma}}, \bibinfo
  {author} {\bibfnamefont {A.}~\bibnamefont {Tang}}, \ and\ \bibinfo {author}
  {\bibfnamefont {Z.}~\bibnamefont {Xu}},\ }\href {\doibase
  10.1016/j.physrep.2018.07.002} {\bibfield  {journal} {\bibinfo  {journal}
  {Phys. Rept.}\ }\textbf {\bibinfo {volume} {760}},\ \bibinfo {pages} {1}
  (\bibinfo {year} {2018})},\ \Eprint {http://arxiv.org/abs/1808.09619}
  {arXiv:1808.09619 [nucl-ex]} \BibitemShut {NoStop}%
\bibitem [{\citenamefont {Hattori}\ and\ \citenamefont
  {Huang}(2017)}]{Hattori:2016emy}%
  \BibitemOpen
  \bibfield  {author} {\bibinfo {author} {\bibfnamefont {K.}~\bibnamefont
  {Hattori}}\ and\ \bibinfo {author} {\bibfnamefont {X.-G.}\ \bibnamefont
  {Huang}},\ }\href {\doibase 10.1007/s41365-016-0178-3} {\bibfield  {journal}
  {\bibinfo  {journal} {Nucl. Sci. Tech.}\ }\textbf {\bibinfo {volume} {28}},\
  \bibinfo {pages} {26} (\bibinfo {year} {2017})},\ \Eprint
  {http://arxiv.org/abs/1609.00747} {arXiv:1609.00747 [nucl-th]} \BibitemShut
  {NoStop}%
\bibitem [{\citenamefont {Woods}\ and\ \citenamefont
  {Saxon}(1954)}]{Woods:1954zz}%
  \BibitemOpen
  \bibfield  {author} {\bibinfo {author} {\bibfnamefont {R.~D.}\ \bibnamefont
  {Woods}}\ and\ \bibinfo {author} {\bibfnamefont {D.~S.}\ \bibnamefont
  {Saxon}},\ }\href {\doibase 10.1103/PhysRev.95.577} {\bibfield  {journal}
  {\bibinfo  {journal} {Phys. Rev.}\ }\textbf {\bibinfo {volume} {95}},\
  \bibinfo {pages} {577} (\bibinfo {year} {1954})}\BibitemShut {NoStop}%
\bibitem [{\citenamefont {Baltz}\ \emph {et~al.}(2009)\citenamefont {Baltz},
  \citenamefont {Gorbunov}, \citenamefont {Klein},\ and\ \citenamefont
  {Nystrand}}]{Baltz:2009jk}%
  \BibitemOpen
  \bibfield  {author} {\bibinfo {author} {\bibfnamefont {A.~J.}\ \bibnamefont
  {Baltz}}, \bibinfo {author} {\bibfnamefont {Y.}~\bibnamefont {Gorbunov}},
  \bibinfo {author} {\bibfnamefont {S.~R.}\ \bibnamefont {Klein}}, \ and\
  \bibinfo {author} {\bibfnamefont {J.}~\bibnamefont {Nystrand}},\ }\href
  {\doibase 10.1103/PhysRevC.80.044902} {\bibfield  {journal} {\bibinfo
  {journal} {Phys. Rev. C}\ }\textbf {\bibinfo {volume} {80}},\ \bibinfo
  {pages} {044902} (\bibinfo {year} {2009})},\ \Eprint
  {http://arxiv.org/abs/0907.1214} {arXiv:0907.1214 [nucl-ex]} \BibitemShut
  {NoStop}%
\bibitem [{\citenamefont {Broz}\ \emph {et~al.}()\citenamefont {Broz},
  \citenamefont {Contreras},\ and\ \citenamefont
  {Tapia~Takaki}}]{Broz:2019kpl}%
  \BibitemOpen
  \bibfield  {author} {\bibinfo {author} {\bibfnamefont {M.}~\bibnamefont
  {Broz}}, \bibinfo {author} {\bibfnamefont {J.~G.}\ \bibnamefont {Contreras}},
  \ and\ \bibinfo {author} {\bibfnamefont {J.~D.}\ \bibnamefont
  {Tapia~Takaki}},\ }\href@noop {} {\ }\BibitemShut {NoStop}%
\bibitem [{\citenamefont {Chiu}\ \emph {et~al.}(2002)\citenamefont {Chiu},
  \citenamefont {Denisov}, \citenamefont {Garcia}, \citenamefont {Katzy},
  \citenamefont {Makeev}, \citenamefont {Murray},\ and\ \citenamefont
  {White}}]{Chiu:2001ij}%
  \BibitemOpen
  \bibfield  {author} {\bibinfo {author} {\bibfnamefont {M.}~\bibnamefont
  {Chiu}}, \bibinfo {author} {\bibfnamefont {A.}~\bibnamefont {Denisov}},
  \bibinfo {author} {\bibfnamefont {E.}~\bibnamefont {Garcia}}, \bibinfo
  {author} {\bibfnamefont {J.}~\bibnamefont {Katzy}}, \bibinfo {author}
  {\bibfnamefont {A.}~\bibnamefont {Makeev}}, \bibinfo {author} {\bibfnamefont
  {M.}~\bibnamefont {Murray}}, \ and\ \bibinfo {author} {\bibfnamefont
  {S.}~\bibnamefont {White}},\ }\href {\doibase 10.1103/PhysRevLett.89.012302}
  {\bibfield  {journal} {\bibinfo  {journal} {Phys. Rev. Lett.}\ }\textbf
  {\bibinfo {volume} {89}},\ \bibinfo {pages} {012302} (\bibinfo {year}
  {2002})},\ \Eprint {http://arxiv.org/abs/nucl-ex/0109018}
  {arXiv:nucl-ex/0109018} \BibitemShut {NoStop}%
\bibitem [{\citenamefont {Abelev}\ \emph {et~al.}(2008)\citenamefont {Abelev}
  \emph {et~al.}}]{STAR:2007elq}%
  \BibitemOpen
  \bibfield  {author} {\bibinfo {author} {\bibfnamefont {B.~I.}\ \bibnamefont
  {Abelev}} \emph {et~al.} (\bibinfo {collaboration} {STAR}),\ }\href {\doibase
  10.1103/PhysRevC.77.034910} {\bibfield  {journal} {\bibinfo  {journal} {Phys.
  Rev. C}\ }\textbf {\bibinfo {volume} {77}},\ \bibinfo {pages} {034910}
  (\bibinfo {year} {2008})},\ \Eprint {http://arxiv.org/abs/0712.3320}
  {arXiv:0712.3320 [nucl-ex]} \BibitemShut {NoStop}%
\bibitem [{\citenamefont {Nakamura}\ \emph {et~al.}(1994)\citenamefont
  {Nakamura} \emph {et~al.}}]{Nakamura:1994zz}%
  \BibitemOpen
  \bibfield  {author} {\bibinfo {author} {\bibfnamefont {T.}~\bibnamefont
  {Nakamura}} \emph {et~al.},\ }\href {\doibase 10.1016/0370-2693(94)91055-3}
  {\bibfield  {journal} {\bibinfo  {journal} {Phys. Lett. B}\ }\textbf
  {\bibinfo {volume} {331}},\ \bibinfo {pages} {296} (\bibinfo {year}
  {1994})}\BibitemShut {NoStop}%
\bibitem [{\citenamefont {Baur}\ \emph {et~al.}(2003)\citenamefont {Baur},
  \citenamefont {Hencken}, \citenamefont {Aste}, \citenamefont {Trautmann},\
  and\ \citenamefont {Klein}}]{Baur:2003ar}%
  \BibitemOpen
  \bibfield  {author} {\bibinfo {author} {\bibfnamefont {G.}~\bibnamefont
  {Baur}}, \bibinfo {author} {\bibfnamefont {K.}~\bibnamefont {Hencken}},
  \bibinfo {author} {\bibfnamefont {A.}~\bibnamefont {Aste}}, \bibinfo {author}
  {\bibfnamefont {D.}~\bibnamefont {Trautmann}}, \ and\ \bibinfo {author}
  {\bibfnamefont {S.~R.}\ \bibnamefont {Klein}},\ }\href {\doibase
  10.1016/j.nuclphysa.2003.09.006} {\bibfield  {journal} {\bibinfo  {journal}
  {Nucl. Phys. A}\ }\textbf {\bibinfo {volume} {729}},\ \bibinfo {pages} {787}
  (\bibinfo {year} {2003})},\ \Eprint {http://arxiv.org/abs/nucl-th/0307031}
  {arXiv:nucl-th/0307031} \BibitemShut {NoStop}%
\bibitem [{STA()}]{STAR:ZDCsmd2003}%
  \BibitemOpen
  \href@noop {} {}\bibinfo {note} {STAR ZDCsmd note,
  \url{https://drupal.star.bnl.gov/STAR/files/ZDC-SMD.pdf}}\BibitemShut
  {NoStop}%
\bibitem [{\citenamefont {Tagliabue}\ and\ \citenamefont
  {Goldemberg}(1961)}]{TAGLIABUE1961144}%
  \BibitemOpen
  \bibfield  {author} {\bibinfo {author} {\bibfnamefont {F.}~\bibnamefont
  {Tagliabue}}\ and\ \bibinfo {author} {\bibfnamefont {J.}~\bibnamefont
  {Goldemberg}},\ }\href {\doibase
  https://doi.org/10.1016/0029-5582(61)90247-4} {\bibfield  {journal} {\bibinfo
   {journal} {Nuclear Physics}\ }\textbf {\bibinfo {volume} {23}},\ \bibinfo
  {pages} {144} (\bibinfo {year} {1961})}\BibitemShut {NoStop}%
\bibitem [{\citenamefont {Bakhtiari}\ \emph {et~al.}(2022)\citenamefont
  {Bakhtiari}, \citenamefont {Jung},\ and\ \citenamefont
  {Lee}}]{Bakhtiari:2022cbn}%
  \BibitemOpen
  \bibfield  {author} {\bibinfo {author} {\bibfnamefont {M.}~\bibnamefont
  {Bakhtiari}}, \bibinfo {author} {\bibfnamefont {N.-S.}\ \bibnamefont {Jung}},
  \ and\ \bibinfo {author} {\bibfnamefont {H.-S.}\ \bibnamefont {Lee}},\ }\href
  {\doibase 10.1016/j.nimb.2022.03.007} {\bibfield  {journal} {\bibinfo
  {journal} {Nucl. Instrum. Meth. B}\ }\textbf {\bibinfo {volume} {521}},\
  \bibinfo {pages} {38} (\bibinfo {year} {2022})}\BibitemShut {NoStop}%
\bibitem [{\citenamefont {Courant}(1951)}]{Courant:1951zz}%
  \BibitemOpen
  \bibfield  {author} {\bibinfo {author} {\bibfnamefont {E.~D.}\ \bibnamefont
  {Courant}},\ }\href {\doibase 10.1103/PhysRev.82.703} {\bibfield  {journal}
  {\bibinfo  {journal} {Phys. Rev.}\ }\textbf {\bibinfo {volume} {82}},\
  \bibinfo {pages} {703} (\bibinfo {year} {1951})}\BibitemShut {NoStop}%
\bibitem [{\citenamefont {Chadwick}\ \emph {et~al.}(2011)\citenamefont
  {Chadwick} \emph {et~al.}}]{Chadwick:2011xwu}%
  \BibitemOpen
  \bibfield  {author} {\bibinfo {author} {\bibfnamefont {M.~B.}\ \bibnamefont
  {Chadwick}} \emph {et~al.},\ }\href {\doibase 10.1016/j.nds.2011.11.002}
  {\bibfield  {journal} {\bibinfo  {journal} {Nucl. Data Sheets}\ }\textbf
  {\bibinfo {volume} {112}},\ \bibinfo {pages} {2887} (\bibinfo {year}
  {2011})}\BibitemShut {NoStop}%
\bibitem [{\citenamefont {Poskanzer}\ and\ \citenamefont
  {Voloshin}(1998)}]{Poskanzer:1998yz}%
  \BibitemOpen
  \bibfield  {author} {\bibinfo {author} {\bibfnamefont {A.~M.}\ \bibnamefont
  {Poskanzer}}\ and\ \bibinfo {author} {\bibfnamefont {S.}~\bibnamefont
  {Voloshin}},\ }\href {\doibase 10.1103/PhysRevC.58.1671} {\bibfield
  {journal} {\bibinfo  {journal} {Phys.Rev.}\ }\textbf {\bibinfo {volume}
  {C58}},\ \bibinfo {pages} {1671} (\bibinfo {year} {1998})}\BibitemShut
  {NoStop}%
\bibitem [{\citenamefont {Xiao}\ \emph {et~al.}(2020)\citenamefont {Xiao},
  \citenamefont {Yuan},\ and\ \citenamefont {Zhou}}]{Xiao:2020ddm}%
  \BibitemOpen
  \bibfield  {author} {\bibinfo {author} {\bibfnamefont {B.-W.}\ \bibnamefont
  {Xiao}}, \bibinfo {author} {\bibfnamefont {F.}~\bibnamefont {Yuan}}, \ and\
  \bibinfo {author} {\bibfnamefont {J.}~\bibnamefont {Zhou}},\ }\href {\doibase
  10.1103/PhysRevLett.125.232301} {\bibfield  {journal} {\bibinfo  {journal}
  {Phys. Rev. Lett.}\ }\textbf {\bibinfo {volume} {125}},\ \bibinfo {pages}
  {232301} (\bibinfo {year} {2020})},\ \Eprint
  {http://arxiv.org/abs/2003.06352} {arXiv:2003.06352 [hep-ph]} \BibitemShut
  {NoStop}%
\bibitem [{\citenamefont {Xing}\ \emph {et~al.}(2020)\citenamefont {Xing},
  \citenamefont {Zhang}, \citenamefont {Zhou},\ and\ \citenamefont
  {Zhou}}]{Xing:2020hwh}%
  \BibitemOpen
  \bibfield  {author} {\bibinfo {author} {\bibfnamefont {H.}~\bibnamefont
  {Xing}}, \bibinfo {author} {\bibfnamefont {C.}~\bibnamefont {Zhang}},
  \bibinfo {author} {\bibfnamefont {J.}~\bibnamefont {Zhou}}, \ and\ \bibinfo
  {author} {\bibfnamefont {Y.-J.}\ \bibnamefont {Zhou}},\ }\href {\doibase
  10.1007/JHEP10(2020)064} {\bibfield  {journal} {\bibinfo  {journal} {JHEP}\
  }\textbf {\bibinfo {volume} {10}},\ \bibinfo {pages} {064} (\bibinfo {year}
  {2020})},\ \Eprint {http://arxiv.org/abs/2006.06206} {arXiv:2006.06206
  [hep-ph]} \BibitemShut {NoStop}%
\bibitem [{\citenamefont {Zhang}\ \emph {et~al.}(2020)\citenamefont {Zhang},
  \citenamefont {Dai},\ and\ \citenamefont {Shao}}]{Zhang:2020onw}%
  \BibitemOpen
  \bibfield  {author} {\bibinfo {author} {\bibfnamefont {C.}~\bibnamefont
  {Zhang}}, \bibinfo {author} {\bibfnamefont {Q.-S.}\ \bibnamefont {Dai}}, \
  and\ \bibinfo {author} {\bibfnamefont {D.~Y.}\ \bibnamefont {Shao}},\ }\href
  {\doibase 10.1007/JHEP02(2023)002} {\bibfield  {journal} {\bibinfo  {journal}
  {JHEP}\ }\textbf {\bibinfo {volume} {23}},\ \bibinfo {pages} {002} (\bibinfo
  {year} {2020})},\ \Eprint {http://arxiv.org/abs/2211.07071} {arXiv:2211.07071
  [hep-ph]} \BibitemShut {NoStop}%
\bibitem [{\citenamefont {Wu}\ \emph {et~al.}(2022)\citenamefont {Wu},
  \citenamefont {Li}, \citenamefont {Tang}, \citenamefont {Wang},\ and\
  \citenamefont {Zha}}]{Wu:2022exl}%
  \BibitemOpen
  \bibfield  {author} {\bibinfo {author} {\bibfnamefont {X.}~\bibnamefont
  {Wu}}, \bibinfo {author} {\bibfnamefont {X.}~\bibnamefont {Li}}, \bibinfo
  {author} {\bibfnamefont {Z.}~\bibnamefont {Tang}}, \bibinfo {author}
  {\bibfnamefont {P.}~\bibnamefont {Wang}}, \ and\ \bibinfo {author}
  {\bibfnamefont {W.}~\bibnamefont {Zha}},\ }\href {\doibase
  10.1103/PhysRevResearch.4.L042048} {\bibfield  {journal} {\bibinfo  {journal}
  {Phys. Rev. Res.}\ }\textbf {\bibinfo {volume} {4}},\ \bibinfo {pages}
  {L042048} (\bibinfo {year} {2022})},\ \Eprint
  {http://arxiv.org/abs/2302.10458} {arXiv:2302.10458 [hep-ph]} \BibitemShut
  {NoStop}%
\bibitem [{\citenamefont {Liang}\ and\ \citenamefont
  {Wang}(2005)}]{Liang:2004ph}%
  \BibitemOpen
  \bibfield  {author} {\bibinfo {author} {\bibfnamefont {Z.-T.}\ \bibnamefont
  {Liang}}\ and\ \bibinfo {author} {\bibfnamefont {X.-N.}\ \bibnamefont
  {Wang}},\ }\href {\doibase 10.1103/PhysRevLett.94.102301} {\bibfield
  {journal} {\bibinfo  {journal} {Phys. Rev. Lett.}\ }\textbf {\bibinfo
  {volume} {94}},\ \bibinfo {pages} {102301} (\bibinfo {year} {2005})},\
  \bibinfo {note} {[Erratum: Phys.Rev.Lett. 96, 039901 (2006)]},\ \Eprint
  {http://arxiv.org/abs/nucl-th/0410079} {arXiv:nucl-th/0410079} \BibitemShut
  {NoStop}%
\bibitem [{\citenamefont {Wang}(2023)}]{Wang:2023fvy}%
  \BibitemOpen
  \bibfield  {author} {\bibinfo {author} {\bibfnamefont {X.-N.}\ \bibnamefont
  {Wang}},\ }\href {\doibase 10.1007/s41365-023-01166-7} {\bibfield  {journal}
  {\bibinfo  {journal} {Nucl. Sci. Tech.}\ }\textbf {\bibinfo {volume} {34}},\
  \bibinfo {pages} {15} (\bibinfo {year} {2023})},\ \Eprint
  {http://arxiv.org/abs/2302.00701} {arXiv:2302.00701 [nucl-th]} \BibitemShut
  {NoStop}%
\bibitem [{\citenamefont {Chen}\ \emph {et~al.}(2023)\citenamefont {Chen},
  \citenamefont {Liang}, \citenamefont {Ma},\ and\ \citenamefont
  {Wang}}]{Chen:2023hnb}%
  \BibitemOpen
  \bibfield  {author} {\bibinfo {author} {\bibfnamefont {J.}~\bibnamefont
  {Chen}}, \bibinfo {author} {\bibfnamefont {Z.-T.}\ \bibnamefont {Liang}},
  \bibinfo {author} {\bibfnamefont {Y.-G.}\ \bibnamefont {Ma}}, \ and\ \bibinfo
  {author} {\bibfnamefont {Q.}~\bibnamefont {Wang}},\ }\href {\doibase
  10.1016/j.scib.2023.04.001} {\bibfield  {journal} {\bibinfo  {journal} {Sci.
  Bull.}\ }\textbf {\bibinfo {volume} {68}},\ \bibinfo {pages} {874} (\bibinfo
  {year} {2023})},\ \Eprint {http://arxiv.org/abs/2305.09114} {arXiv:2305.09114
  [nucl-th]} \BibitemShut {NoStop}%
\bibitem [{\citenamefont {Adler}\ \emph {et~al.}(2001)\citenamefont {Adler},
  \citenamefont {Denisov}, \citenamefont {Garcia}, \citenamefont {Murray},
  \citenamefont {Strobele},\ and\ \citenamefont {White}}]{Adler:2000bd}%
  \BibitemOpen
  \bibfield  {author} {\bibinfo {author} {\bibfnamefont {C.}~\bibnamefont
  {Adler}}, \bibinfo {author} {\bibfnamefont {A.}~\bibnamefont {Denisov}},
  \bibinfo {author} {\bibfnamefont {E.}~\bibnamefont {Garcia}}, \bibinfo
  {author} {\bibfnamefont {M.~J.}\ \bibnamefont {Murray}}, \bibinfo {author}
  {\bibfnamefont {H.}~\bibnamefont {Strobele}}, \ and\ \bibinfo {author}
  {\bibfnamefont {S.~N.}\ \bibnamefont {White}},\ }\href {\doibase
  10.1016/S0168-9002(01)00627-1} {\bibfield  {journal} {\bibinfo  {journal}
  {Nucl. Instrum. Meth.}\ }\textbf {\bibinfo {volume} {A470}},\ \bibinfo
  {pages} {488} (\bibinfo {year} {2001})}\BibitemShut {NoStop}%
\bibitem [{\citenamefont {Baltz}\ \emph {et~al.}(1996)\citenamefont {Baltz},
  \citenamefont {Rhoades-Brown},\ and\ \citenamefont {Weneser}}]{Baltz:1996as}%
  \BibitemOpen
  \bibfield  {author} {\bibinfo {author} {\bibfnamefont {A.~J.}\ \bibnamefont
  {Baltz}}, \bibinfo {author} {\bibfnamefont {M.~J.}\ \bibnamefont
  {Rhoades-Brown}}, \ and\ \bibinfo {author} {\bibfnamefont {J.}~\bibnamefont
  {Weneser}},\ }\href {\doibase 10.1103/PhysRevE.54.4233} {\bibfield  {journal}
  {\bibinfo  {journal} {Phys. Rev. E}\ }\textbf {\bibinfo {volume} {54}},\
  \bibinfo {pages} {4233} (\bibinfo {year} {1996})}\BibitemShut {NoStop}%
\bibitem [{\citenamefont {Harland-Lang}(2023)}]{Harland-Lang:2023ohq}%
  \BibitemOpen
  \bibfield  {author} {\bibinfo {author} {\bibfnamefont {L.~A.}\ \bibnamefont
  {Harland-Lang}},\ }\href {\doibase 10.1103/PhysRevD.107.093004} {\bibfield
  {journal} {\bibinfo  {journal} {Phys. Rev. D}\ }\textbf {\bibinfo {volume}
  {107}},\ \bibinfo {pages} {093004} (\bibinfo {year} {2023})},\ \Eprint
  {http://arxiv.org/abs/2303.04826} {arXiv:2303.04826 [hep-ph]} \BibitemShut
  {NoStop}%
\bibitem [{\citenamefont {Krauss}\ \emph {et~al.}(1997)\citenamefont {Krauss},
  \citenamefont {Greiner},\ and\ \citenamefont {Soff}}]{Krauss:1997vr}%
  \BibitemOpen
  \bibfield  {author} {\bibinfo {author} {\bibfnamefont {F.}~\bibnamefont
  {Krauss}}, \bibinfo {author} {\bibfnamefont {M.}~\bibnamefont {Greiner}}, \
  and\ \bibinfo {author} {\bibfnamefont {G.}~\bibnamefont {Soff}},\ }\href
  {\doibase 10.1016/S0146-6410(97)00049-5} {\bibfield  {journal} {\bibinfo
  {journal} {Prog. Part. Nucl. Phys.}\ }\textbf {\bibinfo {volume} {39}},\
  \bibinfo {pages} {503} (\bibinfo {year} {1997})}\BibitemShut {NoStop}%
\end{thebibliography}%

\end{document}